# Large-Signal Grid-Synchronization Stability Analysis of PLL-based VSCs Using Lyapunov's Direct Method

Yu Zhang, *Student Member, IEEE*, Chen Zhang, and Xu Cai, *Member, IEEE*

*Abstract*—Grid-synchronization stability (GSS) is an emerging stability issue of grid-tied voltage source converters (VSCs), which can be provoked by severe grid voltage sags. Although a qualitative understanding of the mechanism behind the loss of synchronization has been acquired in recent studies, an analytical method for quantitative assessment of GSS of grid-tied VSCs is still missing. To bridge this gap, a dedicated Lyapunov function is analytically proposed, and its corresponding stability criterion for GSS analysis of grid-tied VSCs is rigorously constructed. Both theoretical analysis and simulation result demonstrate that the proposed method can provide a credible GSS evaluation compared to the previous EAC/EF-based method. Therefore, it can be applied for fast GSS evaluation of the grid-tied VSCs as is exemplified in this paper, as well as an analytical tool for some GSS-related issues, e.g., GSS-oriented parameter design and stabilization control.

## I. INTRODUCTION

Recent research of Voltage Source Converters (VSCs) has shown that the grid-tied VSCs are prone to be small-signal unstable under weak grids due to the interactions between VSC control and the grid, in which the dynamics of phase-locked-loop (PLL) plays a vital role. Apart from this, another type of unstable phenomenon relevant to PLL that typically arises from large-signal disturbance (e.g., severe grid fault) and in the form of loss of synchronization (LOS) [1] is emerging, usually referred to as the grid-synchronization stability (GSS) [2].

In this regard, a few studies ranging from the existence of the equilibrium points to the transient stability analysis of the PLL during fault condition have been reported. Thanks to the mathematical similarity between the PLL dynamics and the classic swing equation of a synchronous generator (SG), understanding the LOS of grid-tied VSCs can be greatly assisted with the prior theory of SGs, e.g., the equal-area criterion (EAC) [3, 4] or energy function (EF) [5], which are actually based on Lyapunov's direct method. Although these methods are heuristic in revealing the mechanism of LOS, they are not appropriate for rigorous GSS assessment, as the *indefinite* damping effect (caused by the nonlinearity of the PLL dynamics) has to be ignored to study analytically, leading to an inaccurate region of attraction (ROA) estimation. The existence of the indefinite damping term poses a great challenge to apply direct method. To alleviate this issue, the effect of indefinite damping is taken in to account in [6] by defining the dissipative region, where the damping is positive, but the proposed ROA estimation method is much conservative. Besides, a new Lyapunov function (LF) is proposed in [7], but is only valid under a special case; a numerical LF is computed by sum-of-square programming in [8], whereas the numerical method usually lacks insights into the dynamic essence, which is not helpful to in-depth understanding of the GSS problem.

Therefore, a more practical and accurate way to analytically study the large-signal GSS of the grid-tied VSC by Lyapunov's direct method is still desired. To this end, a further step is taken by proposing an appropriate LF and rigorously developing its corresponding stability criterion for credible GSS assessment of the grid-tied VSC, which is the main contribution of this paper.

## II. SYSTEM MODELING AND STEADY STATES

### A. The GSS Modelling of the Grid-tied VSC

Fig. 1 shows a typical grid-tied VSC which is synchronized by a PLL. Based on the previous modeling method [3, 5], the model for the GSS analysis (referred to as the GSS model) can be developed and written as:

$$\begin{cases} \dot{\delta} = \Delta\omega \\ \dot{x}_{\text{int}} = k_{\text{i,pll}} U_{\text{sq}} \\ \Delta\omega = k_{\text{p,pll}} U_{\text{sq}} + x_{\text{int}} \\ U_{\text{sq}} = -U_{\text{g}} \sin\delta + \Delta\omega L_{\text{g}} I_{\text{sd}} + (R_{\text{g}} I_{\text{sq}} + X_{\text{g}} I_{\text{sd}}) \end{cases} \quad (1)$$

where $\delta = \theta_{\text{pll}} - \theta_{\text{g}}$ represents the phase angle difference between PLL's $d$ axis and the grid voltage vector, $X_{\text{g}} = \omega_{\text{g}} L_{\text{g}}$ is the equivalent grid reactance at normal frequency. Take $\delta$ and $x_{\text{int}}$ as state variables, equation (1) is reorganized as:

$$\begin{cases} \dot{\delta} = \dfrac{k_{\text{p,pll}}(X_{\text{g}} I_{\text{sd}} + R_{\text{g}} I_{\text{sq}} - U_{\text{g}} \sin\delta)}{1 - k_{\text{p,pll}} L_{\text{g}} I_{\text{sd}}} + \dfrac{x_{\text{int}}}{1 - k_{\text{p,pll}} L_{\text{g}} I_{\text{sd}}} \\ \dot{x}_{\text{int}} = \dfrac{k_{\text{i,pll}}(X_{\text{g}} I_{\text{sd}} + R_{\text{g}} I_{\text{sq}} - U_{\text{g}} \sin\delta)}{1 - k_{\text{p,pll}} L_{\text{g}} I_{\text{sd}}} + \dfrac{k_{\text{i,pll}} L_{\text{g}} I_{\text{sd}} \cdot x_{\text{int}}}{1 - k_{\text{p,pll}} L_{\text{g}} I_{\text{sd}}} \end{cases} \quad (2)$$

Equation (2) seems tricky to deal with due to many parameters. For the sake of simplicity, the nondimensionalized form of (2) is derived in (3) by introducing substitutions of (4)(5), where all the variables and parameters are dimensionless quantities without any physical units, including time.

$$\begin{cases} \dfrac{d\delta}{d\tau} = \gamma(m - \sin\delta) + x \\ \dfrac{dx}{d\tau} = (m - \sin\delta) + hx \end{cases} \quad (3)$$

Yu Zhang and Xu Cai are with Laboratory of Control of Power Transmission and Conversion, Shanghai Jiao Tong University, Ministry of Education, Shanghai, China (email: zhangyu666@sjtu.edu.cn, xucai@sjtu.edu.cn).

Cheng Zhang is with Department of Electrical Engineering, Technical University of Denmark (email: chezh@elektro.dtu.dk)

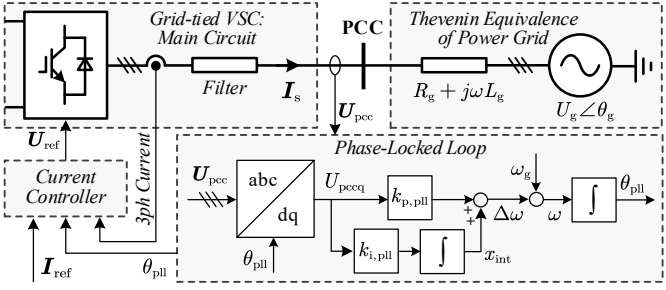

Fig. 1 Schematic diagram of the grid-tied VSC.

$$\gamma = \frac{k_{p,n}\sqrt{\overline{U}_g}}{\sqrt{k_{i,n}}}, \; h = \frac{\sqrt{k_{i,n}}}{\omega_g} \frac{\overline{X}_g \overline{I}_{sd}}{\sqrt{\overline{U}_g}}, \; m = \frac{\overline{X}_g \overline{I}_{sd} + \overline{R}_g \overline{I}_{sq}}{\overline{U}_g} \quad (4)$$

$$x = \frac{x_{int}}{\sqrt{k_{i,n}\overline{U}_g}}, \quad \tau = \frac{\sqrt{k_{i,n}\overline{U}_g} \cdot t}{(1-\gamma h)} \quad (5)$$

In (4)(5), $k_{p,n} = k_{p,pll} U_n, k_{i,n} = k_{i,pll} U_n$ are normalized gains of the PI controller, $U_n$ is the normal grid voltage, and an electrical variable with a head bar represents its per-unit value. It is clear that the dynamics of (3) is governed by three dimensionless parameters: $m, \gamma, h$, which is much easier to analyze compared to equation (2). Besides, it is concluded that:

i. $\gamma > 0$ because PI controller gains are always positive;
ii. $h > 0$ when the VSC is working in the inverter mode, $h = 0$ for STATCOM and $h < 0$ for rectifier;
iii. The proportional gain of PLL's PI controller can be calculated by $k_{p,n} = \alpha_{pll}$, where $\alpha_{pll}$ is PLL's bandwidth which is normally less than $\omega_g$, and $\overline{X}_g \overline{I}_{sd} < 1$ always holds. Hence, $\gamma h = k_{p,n} \overline{X}_g \overline{I}_{sd} / \omega_g < 1$ is usually satisfied.

*B. Equilibrium Points of the System*

Stability of equilibrium points (EPs) should be studied before analyzing the large-signal stability. According to (4), it can be obtained that the EPs exists only if $|m| \leq 1$, satisfying:
$$\sin \delta = m, x = 0. \quad (6)$$

Next, by calculating the eigenvalue of the Jacobian matrix of (3), the condition that the system has stable equilibrium points (SEPs) can be derived as:
$$|m| < 1, h < h_c = \gamma\sqrt{1-m^2}. \quad (7)$$

In this case, the SEPs in (3) are $(2k\pi + \arcsin m, 0)$, $k \in \mathbb{Z}$, and the one in $\delta \in [-\pi, \pi]$ is $(\delta_s, 0)$, where:
$$\delta_s = \arcsin m \quad (8)$$

In summary, (7) is the necessary condition for small-signal stability. However, even if an SEP exists, the system states might hop outside the ROA of the SEP and move toward infinity after large disturbance. Therefore, it is concerned that how much area the SEP attracts, which is studied in the next section.

### III. LYAPUNOV FUNCTION AND STABILITY CRITERION

Enlightened by the EF of SGs and the observation of the indefinite damping of GSS model, the proposed Lyapunov candidate is given as the follows (see Appendix for detail):

$$V(\delta, x) = V_0 + \frac{1}{2}[x - h(\delta - \delta_s)]^2 - (1-\gamma h)(m\delta + \cos\delta), \quad (9)$$

where $V_0 = (1-\gamma h)(m\delta_s + \cos\delta_s)$. The local positiveness of (9)

can be examined by its Hessen matrix, and $(\delta_s, 0)$ is the local minima. The derivative of the proposed LF to time is:

$$\frac{dV}{d\tau} = -\frac{1-\gamma h}{\gamma} \cdot g_1(\delta) \cdot g_2(\delta)$$
$$where: \begin{cases} g_1(\delta) = \gamma(m - \sin\delta) \\ g_2(\delta) = \gamma(m - \sin\delta) + h(\delta - \delta_s). \end{cases} \quad (10)$$

Referring to Fig. 2, if $0 < h < h_c$, then $g_1(\delta) \cdot g_2(\delta) \geq 0$ holds on $\delta \in (\delta_{cr,f}, \delta_{cr,c})$, where $\delta_{cr,c}$ and $\delta_{cr,f}$ are the closet and further zero point of $g_2(\delta)$ near $\delta_s$, respectively:
$$g_2(\delta_{cr,c}) = g_2(\delta_{cr,f}) = 0, \quad (11)$$

indicating that $dV/d\tau \leq 0$ in this region, where it is referred to as the *dissipative region* in this paper. Likewise, the dissipative region for $h \leq 0$ is $\delta \in (\delta_{u,f}, \delta_{u,c})$, where $\delta_{u,c}$ and $\delta_{u,f}$ are the closest and further zero point of $g_1(\delta)$ nearby $\delta_s$, satisfying:
$$g_1(\delta_{u,c}) = g_1(\delta_{u,f}) = 0. \quad (12)$$

Once the dissipative region is determined, the next step is to find a level set, within which $V(\delta, x)$ obeys the same condition as a LF so that the closed region could be an estimated ROA. Intuitively, the area of the estimation should be as large as possible, and the boundary of the largest estimation is the critical level set. To this end, the critical level set should not only be tangent to the boundary of the dissipative region, but also exclude any saddle point (SP) adjacent to the SEP. As is illustrated in Fig. 3, by setting the gradient at the tangent point vertically, the condition that tangent point satisfies is derived as:

$$\frac{d\delta}{dx} = 0 \Rightarrow \frac{\partial V(\delta, x)}{\partial x} = 0 \Rightarrow x = h(\delta - \delta_s), \quad (13)$$

indicating $(\delta_{cr,c}, h(\delta_{cr,c} - \delta_s))$ and $(\delta_{u,c}, h(\delta_{u,c} - \delta_s))$ are the tangent points of the critical level set for $0 < h < h_c$ and $h \leq 0$, respectively. Besides, it can be verified that the value of $V(\delta, x)$ at both tangent points are not greater than that of the closest SP (see Fig. 3 for intuitive illustration), indicating that the level set at tangent points do not enclose any SP. Therefore, the value of the critical level set is:

$$V_{cr} = \begin{cases} V_0 - (1-\gamma h)(m\delta_{cr,c} + \cos\delta_{cr,c}) & \text{if: } 0 < h < h_c \\ V_0 - (1-\gamma h)(m\delta_{u,c} + \cos\delta_{u,c}) & \text{if: } h \leq 0 \end{cases} \quad (14)$$

Then, the large-signal grid-synchronization stability criterion of the grid-tied VSCs comes out as the follows, which can be proved by LaSalle's invariance principle:

**Theorem:** *Suppose $|m| < 1$, $\gamma > 0$, $|\gamma h| < 1$ and $h < \gamma\sqrt{1-m^2}$. The system governed by (3) with initial state:*

$$(\delta(\tau_0), x(\tau_0)): \begin{cases} \delta(\tau_0) \in [\delta_{u,f}, \delta_{u,c}], & \text{if: } 0 \leq m < 1 \\ \delta(\tau_0) \in [\delta_{u,c}, \delta_{u,f}], & \text{if: } -1 < m < 0 \end{cases} \quad (15)$$

*is asymptotically stable if:*
$$V(\delta(\tau_0), x(\tau_0)) \leq V_{cr}. \quad (16)$$

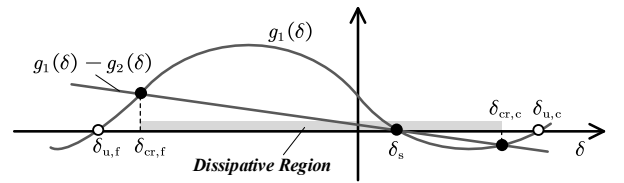

Fig. 2 Illustration of the dissipative region.

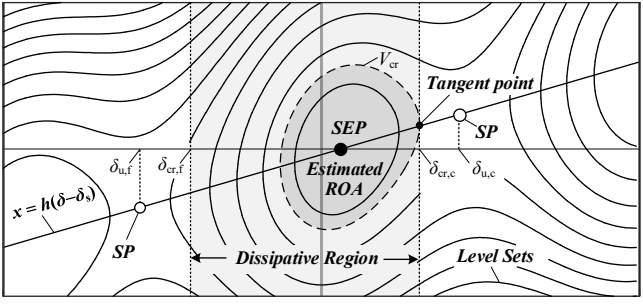

Fig. 3 Critical level set and estimated ROA on ($\delta$, $x$) plane when $0 < h < h_c$.

This theorem actually defines the boundary of the estimated ROA of the SEP, i.e., the critical level set, within which all the initial states are asymptotically stable. Therefore, it provides a sufficient condition for a grid-tied VSC to remain synchronized after the occurrence of a large disturbance.

## IV. VERIFICATION AND AN APPLICATION EXAMPLE

### A. Comparison of the ROA Estimation by Different Methods

A test case with SCR = 2, $\bar{I}_{sd} = 1\text{pu}$, $\bar{I}_{sq} = 0\text{pu}$, $k_{p,n} = 20$, $k_{i,n} = 200\text{s}^{-1}$ is studied, where $m = 0.5$, $\gamma = 1.414$, $h = 0.0225$ could be calculated by (4), and the SEP closest to the origin is (0.5236, 0). Then, ROA for this SEP will be estimated by different methods and plotted in the same phase-plane for comparisons. As shown in Fig. 4, the ROA estimated by the proposed LF is strictly within the real ROA, whereas the ROA estimated by the previous EAC/EF is overoptimistic in some regions (see *Mistakes* in Fig. 4), for it improperly ignores the indefinite damping term[3-5], to cause a part of its estimation exceeds the real boundary. Moreover, it could also be observed that the proposed ROA estimation has the similar size as the EAC/EF-based one. Therefore, the main superiority of the proposed LF compared to the EAC/EF-based method lies in a credible GSS assessment with few sacrifices on the conservativeness.

### B. LF Based GSS Assessment: An Application Example

Another merit of the proposed method is that it enables a fast GSS evaluation, by simply comparing the proposed LF value with the critical value. In this regard, a case study will be conducted as follows: First, a grid voltage dip (i.e., $\bar{V}_g$ is set to 0.2pu to emulate a grid fault) is imposed on the GSS model; Then, the value of the LF while the grid voltage is back to nominal is evaluated, which can be fulfilled by the values of ($\delta$, $x$) acquired from numerical simulation of the GSS model; After this, the concluded stability result will be compared with the simulated phase trajectories to check the credibility of the GSS test. In this case study, four fault durations (i.e., the fault clearing time, FCT) are triggered. The resulting trajectories of the GSS model are shown in Fig. 4.

From the trajectories in Fig. 4, it can be seen that except FCT at $t_4$, the system can remain stable for $t_1 \sim t_3$. This conclusion coincides with that obtained from the real ROA, where the state at $t_4$ is outside the ROA. Next, the proposed method is examined, where the LF values at $t_1 \sim t_4$ along with the stability conclusions are given in Table I. It can be seen that the stability results are the same as the real ROA except $t_3$, thanks to the conservativeness of Lyapunov's direct method. However, in a practical sys-tem, a conservative stability result is better than the optimistic one (see the result of EAC/EF-based method: it concludes that the system is stable at $t_4$ whereas it is unstable in fact), for an optimistic stability conclusion usually implies no countermeasures have to be taken, which may lead to fatal impacts if the system is indeed unstable. Overall, this example demonstrates the advantage of the proposed method in fast and credible GSS test of grid-tied VSCs (i.e., by a simple check of the value of the LF) compared with previous EAC/EF method, though with inevitable conservativeness.

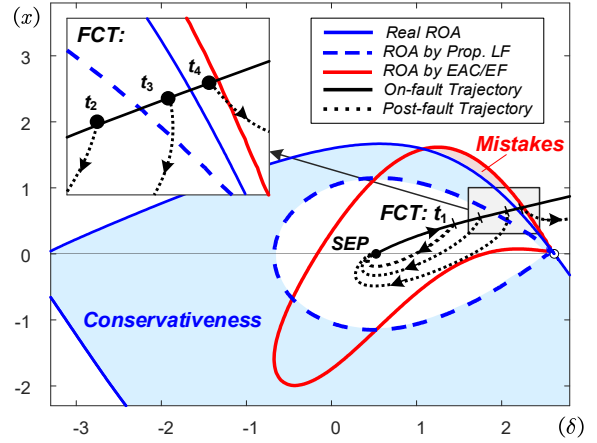

Fig. 4 Comparison of ROA estimation using different methods. The solid blue line denotes the real ROA of the GSS model obtained by computing its stable manifolds from the UEPs. The dotted blue line denotes the estimated ROA with the proposed LF and the critical level set given by (13) (the value is $V_{cr} = 0.6624$ in this case). The red solid line denotes the estimated ROA using the previous EAC/EF-based method. The phase trajectories (line in black) with different FCTs are obtained by simulating the GSS model (3).

TABLE I
STABILITY EVALUATION RESULTS BY DIFFERENT METHODS

| FCT/ms | $V(\delta, x)$ | Evaluation Results | | |
|---|---|---|---|---|
| ($t_1 \sim t_4$) | ($V_{cr} = 0.6624$) | Prop. LF | EAC/EF | Real ROA |
| 80 | 0.2974 | Stable | Stable | Stable |
| 110 | 0.5496 | Stable | Stable | Stable |
| 130 | 0.7318 | Unstable | Stable | Stable |
| 140 | 0.8199 | Unstable | Stable | Unstable |

Moreover, a MATLAB/Simulink simulation is performed with the same circuit and control structure as Fig. 1 (the VSC is in detailed switching model with 200μs controller delay). The results are shown in Fig. 5 and Fig. 6. It can be seen that the PLL frequency $\omega$ would return back normal if the fault is cleared at 80~130ms, while suffer LOS for the 140ms voltage dip, leading to the current beats intermittently. The simulation results comply with the numerical solutions of the GSS model, indicating the feasibility of the GSS model in transient stability assessment of the grid-tied VSCs.

## V. CONCLUSION

This paper derives a novel LF for the PLL and the corresponding ROA estimation method for GSS analysis of the grid-tied VSC based on Lyapunov's direct method. The main contribution of this paper is the successful constructing of a dedicated LF analytically, as well as the developing of a rigorous stability criterion for the large-signal GSS study of the grid-tied VSCs, which, to the best knowledge of the authors, has not been achieved in previous studies. Simulation analysis demonstrates that the proposed ROA is credible compared to the EAC-based method, which can be utilized as a quantitative tool for fast GSS tests. It also lays the foundation for analyzing other relevant issues, such as understanding the parametric influence of GSS, and designing stabilization methods for the LOS. Besides, to the authors' experience, an in-depth understanding of the single-VSC system should be necessary and the first step towards the multi-converter system analysis. Therefore, the proposed LF is believed to inspire the GSS study of multi-VSCs interconnected system through Lyapunov's direct method, which are aspects worthwhile being explored in the future works.

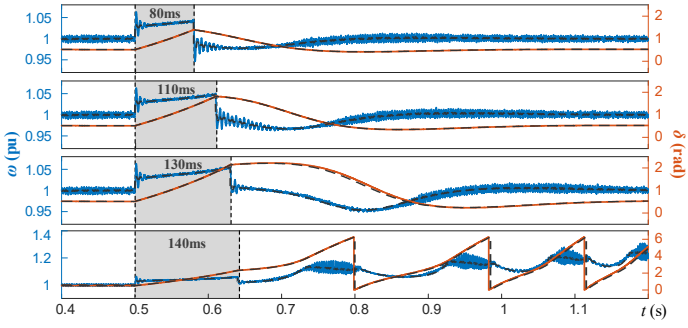

Fig. 5 Simulation results of PLL's frequency $\omega$ and power angle $\delta$ at different FCT (Shaded area: grid voltage drops to 0.2pu, and the time above denotes the period length; Black dashed lines: numerical solutions of the GSS model).

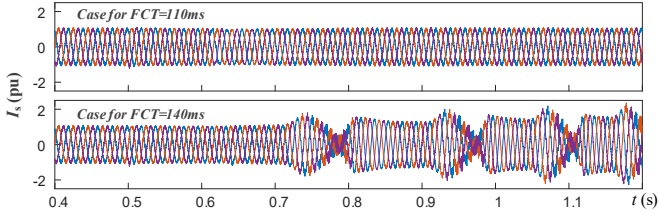

Fig. 6 VSC output current at for stable and unstable scenarios.

## APPENDIX

If $\gamma = h = 0$, the dynamics of the PLL becomes the same as the undamped swing equation of a SG:

$$\frac{d\delta}{d\tau} = x, \quad \frac{dx}{d\tau} = m - \sin\delta \quad (A1)$$

where $x$ represents the rotating frequency, $\delta$ the power angle, and $m$ the driving torque. Referring to its energy function:

$$E(\delta, x) = \frac{1}{2}x^2 + (-m\delta - \cos\delta), \quad (A2)$$

it can be examined that $E(\delta, x)$ is time invariant for (A1). Next, if $\gamma$ and $h$ is considered, take its derivative to time:

$$\frac{dE}{d\tau} = hx^2 - \gamma(m - \sin\delta)^2 \quad (A3)$$

which is negative semidefinite when $h \leq 0$ and indefinite when $h > 0$. Therefore, (A2) can be an EF (a kind of local LF) of (3) when the grid-tied VSC is in the rectifier mode or STATCOM mode, while not capable for the inverter mode. In this regard, modification should be made to get a versatile LF. According to (A3), $\gamma$ can be regarded as the positive damping, while $h$ be the negative damping in the inverter mode. Therefore, endeavors are made to shift some damping effect in (A3) from $-\gamma(m-\sin\delta)^2$ to $hx^2$ by a positive (semi-) definite quadratic form. Supposing the LF of dynamic equation (3) be:

$$V(\delta, x) = V_0 + \frac{1}{2}\mathbf{x}^T \begin{bmatrix} 1 & a \\ a & pa^2 \end{bmatrix} \mathbf{x} - \varepsilon(m\delta + \cos\delta) \quad (A4)$$

where $\mathbf{x} = [x \ \delta - \delta_s]^T$, $a, p, \varepsilon$ are undetermined coefficients, $p \geq 1$ to ensure the quadratic form be positive (semi-)definite. Then, take the derivative of (A4) to time:

$$\begin{aligned}\frac{dV}{d\tau} = &(h+a)x^2 - \varepsilon\gamma(m - \sin\delta)^2 \\ &+ (a + pa^2\gamma)(\delta - \delta_s)(m - \sin\delta) \\ &+ (1 + a\gamma - \varepsilon)(m - \sin\delta)x + (ah + pa^2)(\delta - \delta_s)x\end{aligned} \quad (A5)$$

Let the coefficient of the coupling terms be zero, and that of the square terms be non-positive, leading to the relations below:

$$1 + a\gamma - \varepsilon = 0, \quad ah + pa^2 = 0, \quad h + a \leq 0,$$

The unique solution is $a = -h, p = 1, \varepsilon = 1 - \gamma h$, and $V_0$ is chosen to make sure $V(\delta_s, 0) = 0$. In this way, (A4) is derived as:

$$V(\delta, x) = V_0 + \frac{1}{2}[x - h(\delta - \delta_s)]^2 - (1 - \gamma h)(m\delta + \cos\delta) \quad (A6)$$

which is endowed with the LF properties near the SEP.